\documentclass[letterpaper, 10 pt, conference]{ieeeconf}   

\usepackage{graphicx}
\usepackage{multirow}
\usepackage{float}
\usepackage{diagbox}
\usepackage{cite}
\usepackage{amsmath}

\IEEEoverridecommandlockouts                              

\title{\LARGE \bf
Efficient Hybrid Neuromorphic-Bayesian Model for Olfaction Sensing: Detection and Classification
}

\author{Rizwana Kausar$^{1}$, Fakhreddine Zayer$^{1}$, Jaime Viegas$^{1}$, and Jorge Dias $^{1}$
\thanks{$^{1}$ Center for Autonomous Robotic Systems, Khalifa University, Abu Dhabi, UAE } %
}

\begin{document}

\maketitle
\thispagestyle{empty}
\pagestyle{empty}

\begin{abstract}
Olfaction sensing in autonomous robotics faces challenges in dynamic operations, energy efficiency, and edge processing. It necessitates a machine learning algorithm capable of managing real-world odor interference, ensuring resource efficiency for mobile robotics, and accurately estimating gas features for critical tasks such as odor mapping, localization, and alarm generation. This paper introduces a hybrid approach that exploits neuromorphic computing in combination with probabilistic inference to address these demanding requirements.
Our approach implements a combination of a convolutional spiking neural network for feature extraction and a Bayesian spiking neural network for odor detection and identification. The developed algorithm is rigorously tested on a dataset for sensor drift compensation for robustness evaluation. Additionally, for efficiency evaluation, we compare the energy consumption of our model with a non-spiking machine learning algorithm under identical dataset and operating conditions. Our approach demonstrates superior efficiency alongside comparable accuracy outcomes.
   
\end{abstract}

\section{INTRODUCTION}
\begin{figure}[!ht]
      \centering
      \includegraphics[scale=0.35]{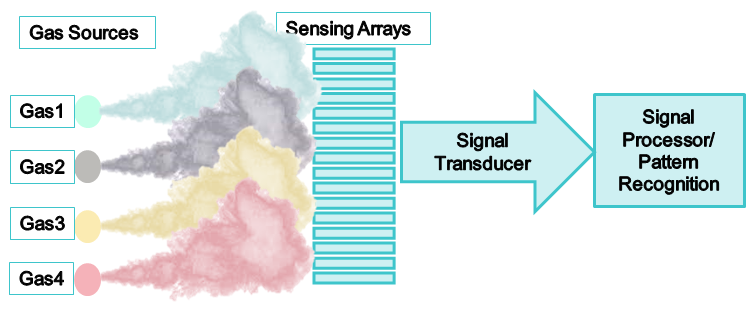}
      \caption{Overview of an Artificial Olfaction System. }
      \label{Olfaction}
   \end{figure}

Integrating a robot and environmental monitoring sensors through a potent machine-learning algorithm not only streamlines process automation but also ensures reliable monitoring. It generates timely alerts well in advance of safety escalation. However, guaranteeing the complete autonomy of such operations necessitates a machine learning algorithm that exhibits energy efficiency, possesses edge processing capabilities, and can operate reliably in challenging environmental conditions \cite{articleML}. One prominent example of such interaction between artificial intelligence and robotic sensors can be observed in olfactory monitoring. Artificial olfaction sensing, often referred to as an "Electronic Nose," has wide-ranging applications, including assessing fruit ripeness in agriculture \cite{articleFruit}, monitoring air quality in environmental contexts \cite{articleEnvironment}, detecting food adulteration in the food industry \cite{article3Food}, identifying drugs and explosives in security applications \cite{bookApplications6}, and facilitating non-invasive early-stage disease discovery in healthcare \cite{articleDesease}. These diverse applications demand specialized electronic noses, where each component is tailored to operate effectively under specific conditions with corresponding AI algorithms. Fig \ref{Olfaction} illustrates a fundamental design structure of an electronic nose, composed of a sensor array module and machine learning-based pattern recognition, interconnected through a readout circuit. 
Unlike color or sound, odors are difficult to standardize and describe consistently. Challenges like variability in environmental conditions, humidity, the presence of other odors, and sensor response degradation can complicate the interpretation of sensor data. Addressing these challenges is paramount to unlocking the full potential of artificial olfaction systems, making them more reliable and versatile across a myriad of practical applications \cite{articleFruit,articleEnvironment,article3Food,bookApplications6,articleDesease}.

 The introduction of artificial neural networks marked a transformative milestone in olfaction sensing, ushering in high detection accuracy, flexible hardware options, and robust algorithms. This progression has significantly advanced the capabilities of electronic noses, making them more adept at addressing real-world challenges in odor detection and classification.  Several machine learning algorithms have been employed to resolve these challenges \cite{articleDatapreprocess7} including Principal Component Analysis (PCA) \cite{6564152PCA}, Support Vector Machine (SVM) \cite{8167170SVM}, Decision Tree (DT) \cite{CHO2011542DT}, and K-nearest neighbor (KNN) \cite{s121115542KNN}. While these algorithms excel in the context of stationary odor analysis, the demands of mobile robotics, which necessitate a combination of autonomy, energy efficiency, and robustness, call for the development of a considerably more stable algorithm.

In this study, we present a cutting-edge novel neuromorphic model architecture. This architecture is specifically engineered to tackle two paramount challenges in olfaction sensing: excessive power consumption and robustness. Our innovation centers on a novel hybrid approach, combining the strengths of Spiking Neural Networks (SNN) for feature extraction and odor concentration estimation seamlessly integrated with a Bayesian-based olfactory cortex for precise odor detection and classification. Overall, our work not only addresses critical issues in olfaction sensing but also pushes the boundaries of what is achievable, offering a powerful and efficient solution with far-reaching implications. The contributions are noteworthy in several ways:
\begin{itemize}
    \item Mitigation of Power Consumption: We address a critical issue in artificial olfaction by substantially reducing power consumption. Through the integration of a spike-based neuromorphic framework, our approach achieves remarkable energy efficiency, outperforming conventional methods.
    \item Enhanced Robustness: By leveraging Bayesian principles within our hybrid architecture, we markedly improve the model's generalization and robustness. This advantage empowers our system to excel in the complex and variable landscape of odor patterns, surpassing the capabilities of existing approaches.
\end{itemize}

This paper follows a structured approach: Section II provides a comprehensive overview of state-of-the-art neuromorphic-based classifiers and Bayesian networks. Section III discusses our neuromorphic classifier approach, addressing challenges such as power consumption and robustness. In Section IV, we present details regarding selected datasets and a few preprocessing steps. Section V presents experimental results and discussions, offering insights and implications. Finally, in Section VI, we draw conclusions based on our contributions, emphasizing the transformative impact of our work in the field of artificial olfaction.

\section{RELATED WORK}
\subsection{Spiking Neural Networks}
Unlike traditional artificial neural networks, which typically use continuous-valued signals, Spiking Neural Networks (SNNs) model the behavior of neurons firing action potentials, also known as spikes. This introduces sparsity in both connectivity and activity patterns, resulting in significantly improved power efficiency compared to standard machine learning algorithms. SNNs have found utility in diverse domains, including artificial olfaction sensing, where they offer distinctive advantages. In the context of olfaction, SNNs leverage their temporal processing capabilities to analyze complex odor patterns over time, making them well-suited for discerning subtle differences in scents \cite{articleLi}. Moreover, SNNs align closely with the spatiotemporal dynamics of biological olfactory systems, enhancing their applicability in mimicking natural olfaction processes \cite{10112635}.
A three-stage approach, mirroring the insect olfactory system as reported in \cite{article18}, is utilized for odor classification. The Spike-Timing-Dependent Plasticity (STDP) based learning model discussed in \cite{articlebased}, coupled with Spike reward-dependent plasticity (SRDP), achieves an accuracy of 94\%. Similarly, \cite{articleInsect} presents a comparable approach for classifying the VOC dataset with an accuracy of 92\%.

In \cite{10112635}, a hybrid model combining SNN and SVM is reported for liquor classification, achieving an accuracy of 93\%. The study in \cite{10112635} models a mammal olfactory system using bionic data processing methods, yielding higher accuracy than conventional methods. It concludes that reciprocal connections provide superior accuracy compared to lateral connections.

Additionally, \cite{articleHam} discusses an online learning mammal olfactory system, performing gas classification based on Hamming distance and employing few-shot learning. An online learning algorithm based on the external plexiform layer (EPL) is detailed in \cite{articleSTDP}, showing improved performance compared to conventional machine learning methods when tested on publicly available datasets.

Furthermore, \cite{s20102756Reservoir} implements a 3D-SNN model using the Java-based NeuCube framework SNN reservoir for odor source classification. Finally, \cite{9789674Audio} designs an SNN model for auditory and olfactory data fusion to achieve odor source location and mapping in mobile robots.


The above-mentioned SNN methods focus solely on classification through online learning with an STDP-based approach. In contrast, our proposed method is a hybrid approach utilizing supervised learning to compensate for sensor drift while conducting odor classification.

In addition to olfaction, SNNs have proven valuable in neuroscience research \cite{inproceedingsnEURO}, robotics \cite{articlerOBOT}, neuromorphic engineering \cite{articleSNNinNeuromorphic}, and pattern recognition tasks involving temporal data, such as audio signal processing \cite{9789674Audio}. However, SNNs pose unique challenges in terms of training algorithms and specialized hardware implementations \cite{articleSNNinNeuromorphic}. Notable software frameworks for SNN simulation and training include NEURON \cite{articleNeuron}, Brian \cite{articleBrian}, SpiNNaker \cite{articleSpinnaker}, Nengo \cite{articleNengo}, SNNTorch \cite{10242251SNNtorch} and hardware-accelerated platforms like NEST \cite{inbookNest} and Intel's Loihi \cite{articleLoihi}.

\subsection{SNN Background Theorey} 
The primary difference between a spiking neural network and an artificial neural network lies in the type of neuron utilized. Unlike traditional artificial neurons in ANNs, which typically compute a continuous output based on weighted inputs and apply a non-linear activation function, spiking neurons operate by generating discrete spikes in response to input stimuli, devoid of any activation function. When the membrane potential $V_{m}$ of the neuron surpasses a threshold $V_{th}$,  a spike is generated, which then propagates to its connected neurons. 
The most common type of spiking neuron is the leaky integrate and fire (LIF) neuron. LIF neuron's potential update and firing mechanism can be modeled using an RC circuits low-pass filter model as given by \cite{10242251SNNtorch}. 
\begin{equation}
 \alpha \frac{dV(t)}{dt}=-V(t)+I_{in}(t)R
\end{equation}
where $\alpha=RC_m$ is a time constant. The approximation of the above equation through the Euler method is given by 
\begin{equation}
V(t)=\gamma V(t-1) +(1-\gamma)I_{in}(t)
\end{equation}
where $\gamma=e^{-\frac{1}{\alpha}}$ is decay rate in the membrane potential of neuron. $I_{in}(t)$ can be taken as the input feature vector $X_{in}[t]$. 
\begin{equation}
 \label{eq:main}
V[t]=\gamma V[t-1] +W(X_{in}[t])- Spk_{out}[t-1]V_{thr}
\end{equation} 
The membrane potential of the LIF neuron at any time 't' is given by the above equation \ref{eq:main}, where the first term on the right-hand side of the equations depicts the decay in the potential, the second term is the weighted input to the neuron and the third term is a soft reset in case of a spike. The condition for spike generation is given below:
\begin{equation}
Spk_{out}[t]=
\begin{cases}
      1, & \text{if}\ V[t]>V_{thr} \\
      0, & \text{otherwise}
    \end{cases}
\end{equation}

The input feature vector, encoded as spikes, influences the synaptic current received by neurons. This synaptic current, in turn, dictates the membrane potential of the neuron. When the membrane potential surpasses a predefined threshold, a spike is generated. In the rate encoding scheme, the classification decision is contingent on the number of spikes produced by a neuron. If a particular neuron generates more spikes than all other neurons, the model outputs the corresponding class.
\subsection{Bayesian Neural networks} 

In contrast to traditional neural networks, which generate deterministic outputs, Bayesian Neural Networks (BNNs) provide a range of possible outcomes. This characteristic helps alleviate overfitting, facilitates learning from limited datasets, and enhances model generalization, enabling effective predictions even on noisy data that falls outside the range of the training distribution.
This capability arises from the probabilistic modeling of weight and bias distributions within the network, allowing it to continually refine its predictions based on prior knowledge \cite{quteprints20200945}. Moreover,  The uncertainty provision by Bayesian in decision-making enables our model to evaluate its output before generating a gas detection alarm, making it a feasible solution for highly critical applications like chemical industry and nuclear plant emission monitoring.
Let 'x' be the training data with 'y' as corresponding training labels and $\alpha$ as a set of weights and biases, we represent this in conventional neural networks as a linear function $y=f(x,\alpha)$ which will provide a fix $\alpha$ for a given value of x and y. However, in the case of the Bayesian Neural Network, the same function becomes a probability distribution decided by all possible outcomes of $P(y|x,X)$, where;
\begin{equation}
 P(y|x,X)=\int (P(y|x, \alpha)*P(\alpha|X) \,d\alpha \
\end{equation}

The first term $(P(y|x, \alpha)$ in the above equation is a likelihood function that estimates the probability of having y for a given value of x and $\alpha$. The second term is called posterior distribution of weights decided based on data X observed by the model. Now to make a prediction we use;

\begin{equation}
 P(\hat{y}|\hat{x},X)=\int (P(\hat{y}|\hat{x}, \alpha)P(\alpha|X) \,d\alpha \
\end{equation}

here $\hat{x}$ is the  new sample and $\hat{y}$ is the predicted output. The major challenge in Bayesian Neural networks is estimation of posterior distribution $P(\alpha|X)$. We can approximate the true posterior with a variational distribution $Q(\alpha|D)$  of a known functional form, for which we aim to estimate the parameters. This approximation involves minimizing the Kullback-Leibler divergence between $Q(\alpha|D)$ and true posterior $P(\alpha|X)$.

Bayesian inference-based gas detection and localization are performed in \cite{article50} through a coarse model approach. \cite{571194451} discusses a Bayesian-based fusion approach for the detection and classification of gas data. \cite{article46} discussed a Bayesian inference-based mammal olfactory system model discussed in detail with reward-based learning techniques. 
All these techniques are non-spiking, standalone methods, whereas we have implemented a spiking Bayesian hybrid approach to address sensor drift.

\section{PROPOSED APPROACH}
\begin{figure}[H]
      \centering

      \includegraphics[scale=0.2]{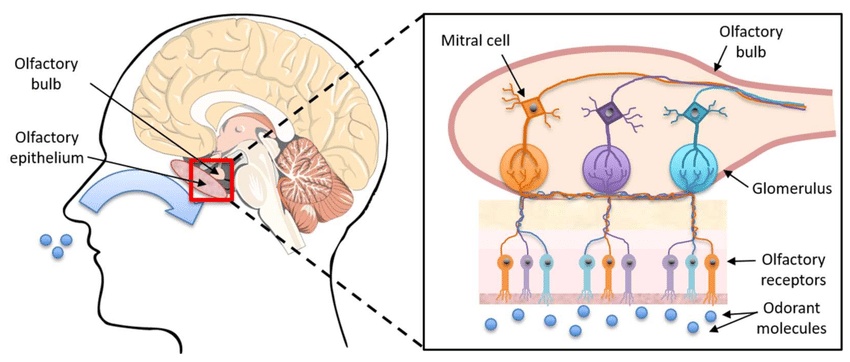}
      \caption{Human Olfactory System \cite{articleolfactory}.}
      \label{Olfactory}
      \vspace{-3mm}
   \end{figure}

To provide insight into the rationale behind the proposed approach, a concise overview of the human olfactory system is presented in the following paragraph.
The human olfactory system, Fig \ref{Olfactory}, consists of a large family (in the order of millions) of olfactory receptors (ORs) where each receptor is sensitive to specific odor molecules. These receptors in interaction with odor molecule that matches their specificity generate signals called nerve impulses (action potential).

These spikes carry information about the odor to the brain. These olfactory nerves (ONs) project the signal to the olfactory bulb (OB). Olfactory nerves transmit the signal to a particular location in OB called the glomerulus. A particular glomerulus receives the signal from several ONs (sensitive to the same odor molecule). In other words, ONs carrying the signals from a particular set of ORs sensitive to a specific odor molecule will end up on the same glomerulus in OB. From glomerulus, odor information is sent to the primary olfactory cortex through mitral cells. In the OB, the incoming signals are processed for odor classification in the olfactory cortex\cite{articleHumanOlf}.

 

    \begin{figure*}[thpb]
      \centering

      \includegraphics[scale=0.45]{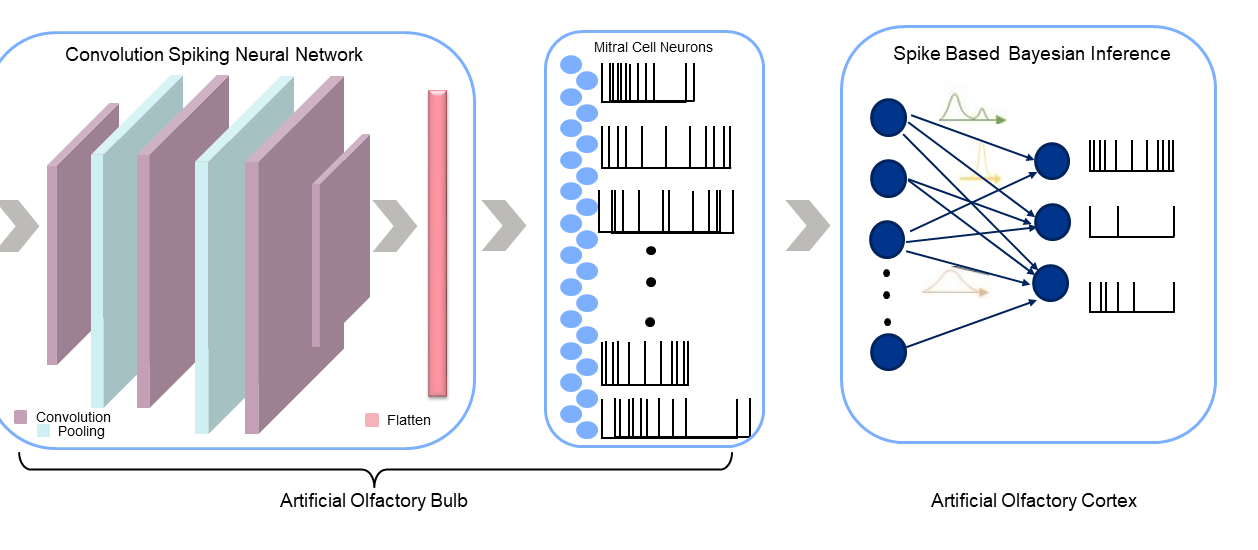}
      \caption{Detailed Approach for Artificial Olfaction Computation.}
      \label{Detail}
   \end{figure*}

\begin{table*}[htbh]
\caption{Comparison of Proposed Model Response with other Reported Architectures for Short-Term Drift }
\label{Short Term}
\begin{center}
\begin{tabular}{|p{2.cm}|p{1cm}p{1cm}p{1cm}p{1cm}p{1cm}p{1cm}p{1cm}p{1cm}p{1cm}p{1.2cm}|}
\hline 
 \diagbox{Model}{Batch}
& 1 to 2 & 2 to 3 & 3 to 4& 4 to 5&5 to 6& 6 to 7&7 to 8& 8 to 9&9 to 10& Avg
\\
\hline
SVM-brf  & 74.36& 87.83&90.06&56.35&42.52&83.53&91.84&62.98&22.64&68.01
\\
SVM-comgfk  & 74.47& 73.75&78.51&64.26&69.97&77.69&82.69&85.53&17.76&69.40
\\
ML-comgfk  & 80.25 & \textbf{98.55}&84.89&89.85&75.53&91.17&61.22&95.53&39.56&79.62
\\
ELM-rbf& 70.63&40.44&64.16&64.37&72.2&80.75&88.2&67&22&63.36
\\
DAELM-S(5)& 72.66&69.99&72.61&79.54&52.93&87.18&91.36&56.66&29.05&68
\\
DTBLS & 78.67&97.65&79.88&67.01&75.34&90.44&\textbf{95.1}&68.09&54.47&78.52
\\
TDACNN&89.56&97.46&87.58&94.68&73.9&80.18&78.43&83.19&47.64&81.48
\\
SWKELM & 76.13&90.73&90.68&93.3&73.43&83.73&87.07&92.55&41.69&81.05
\\
DTSWKELM & 88.26&97.23&\textbf{93.17}&\textbf{98.98}&78.43&\textbf{93.99}&93.19&\textbf{96.38}&55.08&\textbf{88.30}
\\
Neuro-Bayesian & \textbf{94.12} & 92.5 & 93.05&86.7&\textbf{88.5}&83.0&87.5&86.5&\textbf{81.2}& 88.12
\\
\hline
\end{tabular}
\end{center}
\end{table*}

\begin{table*}[h]
\caption{Comparison of Proposed Model Response with other Reported Architectures for Long-Term Drift }
\label{Long Term}
\begin{center}
\begin{tabular}{|p{2.cm}|p{1cm}p{1cm}p{1cm}p{1cm}p{1cm}p{1cm}p{1cm}p{1cm}p{1cm}p{1.2cm}|}
\hline 
 \diagbox{Model}{Batch}

& 1 to 2 & 1 to 3 & 1 to 4& 1 to 5&1 to 6& 1 to 7&1 to 8& 1 to 9&1 to 10& Avg
\\
\hline
SVM-brf  & 74.36& 61.03&50.93&18.27&28.262&28.81&20.07&34.26&34.48&38.94
\\
SVM-comgfk  & 74.47& 70.15&59.78&75.09&73.99&54.59&55.88&70.23&41.85&64
\\
ML-comgfk  & 80.25 & 74.99&78.79&67.41&77.82&71.68&49.96&50.79&53.79&67.28
\\
ELM-rbf& 70.63&66.44&66.83&63.45&69.73&51.23&49.76&49.83&33.5&57.93
\\
DAELM-S(5)& 72.66&75.72&61.3&86.29&53.45&59.4&31.16&66.85&44.39&61.25
\\
DTBLS & 78.67&\textbf{96.36}  &74.6&85.23&83.2&81.53&58.67&56.19&63.1&75.28
\\
TDACNN&89.56&83.83&77.64&75.63&74.36&62.08&\textbf{75.1} &60.85&50.88&72.21
\\
SWKELM & 76.13&86.88&73.29&81.82&89.73&71.88&43.57&59.78&53.31&70.71
\\
DTSWKELM & 88.26&90.66&77.01&89.85&\textbf{96.31} &74.29&55.78&62.77&68.33&78.14
\\
Neuro-Bayesian& \textbf{94.12} & 93.35 & \textbf{87.15}&\textbf{91.21}& 83.11
&\textbf{87.14}
&74.13 &\textbf{76.23}&\textbf{71.4}&\textbf{84.20}
\\
\hline 
\end{tabular}
\end{center}
\end{table*}
   
Our proposed method emulates the functionality of the human olfactory system through a combination of Spiking-based feature extraction and Bayesian-based inference. The input module, responsible for data preprocessing and spike generation, functions as an artificial olfactory receptor (AOR) layer. The artificial olfactory bulb module consists of two layers: a convolution spike layer, akin to glomeruli, and a hidden layer that simulates mitral cells. Subsequently, a Bayesian classifier serves the role of the olfactory cortex. As debated in [33] human olfaction is more like a probabilistic system that processes the information more closely in a Bayesian way than conventional neural network classification techniques. Moreover, in real-world scenarios, systems are often exposed to multiple odors or combinations of odors, alongside environmental noise. In such complex environments, probabilistic inference has demonstrated greater suitability compared to conventional techniques [34].
An input rate-based data encoding scheme is employed for the generation of spikes while Bayesian inference is performed through Kullback-Leibler divergence. Odor category prediction is made through spike count at the output neurons.  

In our novel hybrid approach, spiking-based computation contributes to power efficiency, while Bayesian inference addresses sensor drift by leveraging Bayesian robustness.

\section{Dataset}
The Gas Sensor Array Drift Dataset, available on the UCI Machine Learning online repository, is utilized to evaluate the proposed methodology\cite{misc_gas_sensor_array_drift_dataset_224}.
The dataset spans 36 months and utilizes a 16-metal oxide sensor array. Specifically, four sensors of each type—TGS2600, TGS2602, TGS2610, and TGS2620—manufactured by Figaro Inc. These sensors are exposed to six distinct odorants (1: Ethanol; 2: Ethylene; 3: Ammonia; 4: Acetaldehyde; 5: Acetone; 6: Toluene). The resulting dataset is partitioned into 10 batches as given in Table \ref{table_Dataset}.
Each sensor's response is recorded in the form of eight features, encompassing two steady-state features and six transient features as mentioned in Table \ref{table_Feature}. Steady-state features comprise the maximum change in resistance from the baseline and the normalized value of this change. The remaining six features pertain to the transient characteristics, depicting the ascending (max $ema_\alpha$) and descending (min $ema_\alpha$) trends within the sensor's curve response. These attributes are derived using the exponential moving average  ($ema_\alpha$) transformation, which converts the discrete time series response r[k] into a scalar value \cite{articleGasDrift}. 

\begin{equation}
\Delta R= max_k(r[k])-min_k(r[k])
\end{equation}
\begin{equation}
||\Delta R||=\frac{max_k(r[k])-min_k(r[k])}{min_k(r[k])} 
\end{equation}

\begin{equation}
ema_\alpha[k]= (1-\alpha)ema_\alpha[k-1]+\alpha(r[k]-r[k-1])
\end{equation}

For each instance captured by 16 sensors array, there are $8 \times 16$, i.e 128 feature values. 
\begin{table}[htbh]
\caption{Dataset Feature Set Detail}
\label{table_Feature}
\begin{center}
\begin{tabular}{|p{2.5cm}|p{2cm}|p{2cm}|}
\hline 
Steady State Features
& Transient Feature (Rise) 
$ max_k ema_\alpha(r[k])$&
Transient Features 
(Decay)
$ min_k ema_\alpha(r[k])$
\\
\hline
$\Delta R $ & $ \alpha=0.001$ & $ \alpha=0.001$ 
\\(Max Change in R) & $\alpha=0.01$ & $\alpha=0.01$
\\
$||\Delta R||$           (Normalized Resistance)& $ \alpha=0.1$&$ \alpha=0.1$
\\
\hline
\end{tabular}
\end{center}
\vspace{-5mm}
\end{table}

\begin{table*}[htbh]
\caption{Dataset Detail}
\label{table_Dataset}
\begin{center}
\begin{tabular}{|p{1.05cm}|p{3cm}|p{1cm}|p{1cm}|p{1cm}|p{1cm}|p{1.25cm}|p{1cm}|p{1cm}|}
\hline
Batch &Duration (Month ID) & Total Instances & Ethanol& Ethylene & Ammonia & Acetaldehyde & Acetone & Toluene 
\\
\hline
Batch 1 & Months 1 and 2   
& 445 & 83 & 30 & 70 & 98 & 90 & 74
\\
\hline
Batch 2 & Months 3, 4, 8, 9 and 10 
 & 1244 &  100 & 109 & 532 & 334 & 164 & 5 
\\
\hline
Batch 3 & Months 11, 12, and 13
& 1586 & 216 & 240 & 275 & 490 & 365 & 0
\\
\hline
Batch 4	& Months 14 and 15 & 161 & 12&  30 & 12& 43& 64& 0
\\
\hline
Batch 5 & Month 16 & 197 & 20 & 46&63& 40& 28& 0
\\
\hline
Batch 6 & Months 17, 18, 19, and 20 & 2300 & 110 & 29 & 606 & 574 & 514 & 467
\\
\hline
Batch 7 & Month 21 & 3613 & 360 & 744 & 630 & 662 & 649 & 568
\\
\hline
Batch 8	 & Months 22 and 23 & 294 & 40 & 33 & 143 & 30 & 30 & 18
\\
\hline
Batch 9	& Months 24 and 30 & 470 & 100 & 75 & 78 & 55 & 61 & 101
\\
\hline
Batch 10 & Month 36 & 3600 & 600 & 600 & 600 & 600 & 600 & 600
\\
\hline
\end{tabular}
\end{center}
\vspace{-5mm}
\end{table*}
\begin{figure}[thpb]
      \centering

      \includegraphics[scale=0.20]{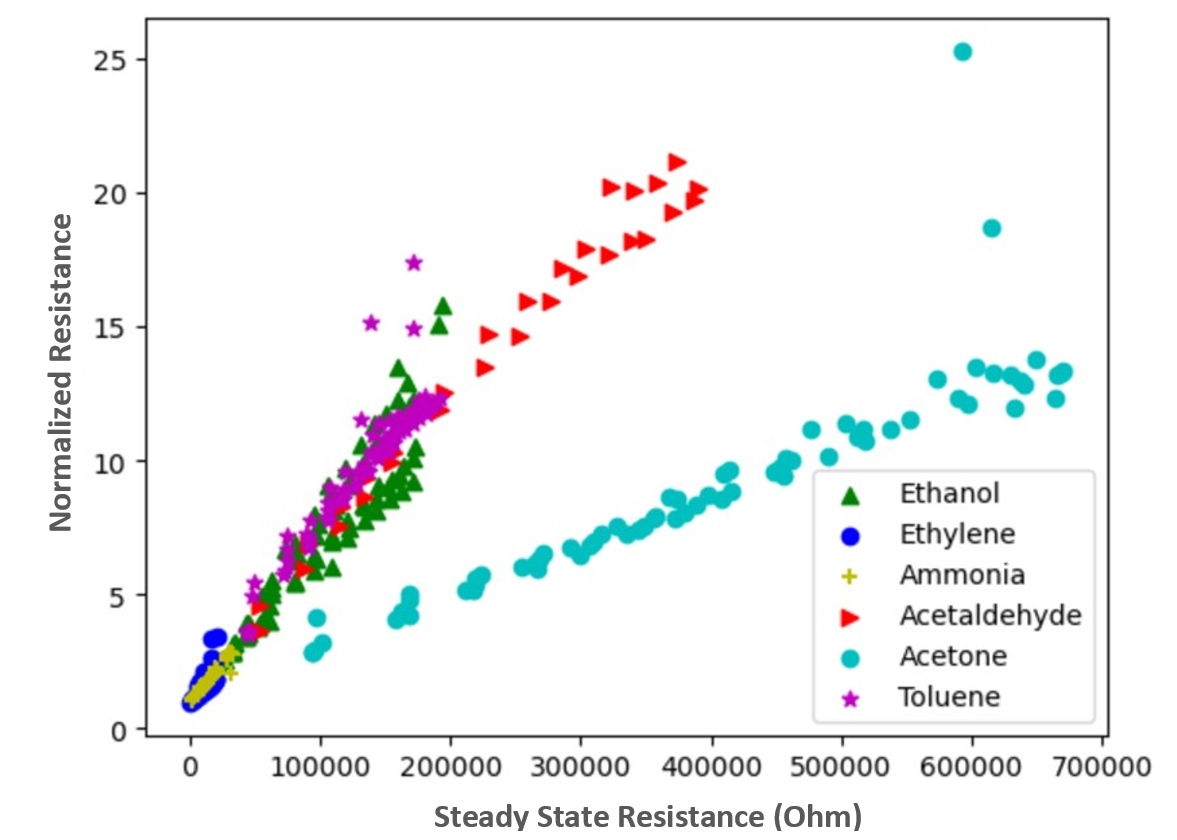}
      \caption{Steady State Features Spread in the selected Dataset.}
      \label{Dataset Spread}
      \vspace{-3mm}
   \end{figure}
The normalized and steady-state feature spread of the dataset for each class is depicted in Fig \ref{Dataset Spread} showing the huge spread difference within the classes. To handle this we have implemented a logarithmic-based data normalization technique.      

\begin{figure}[thpb]
      \centering

      \includegraphics[scale=0.25]{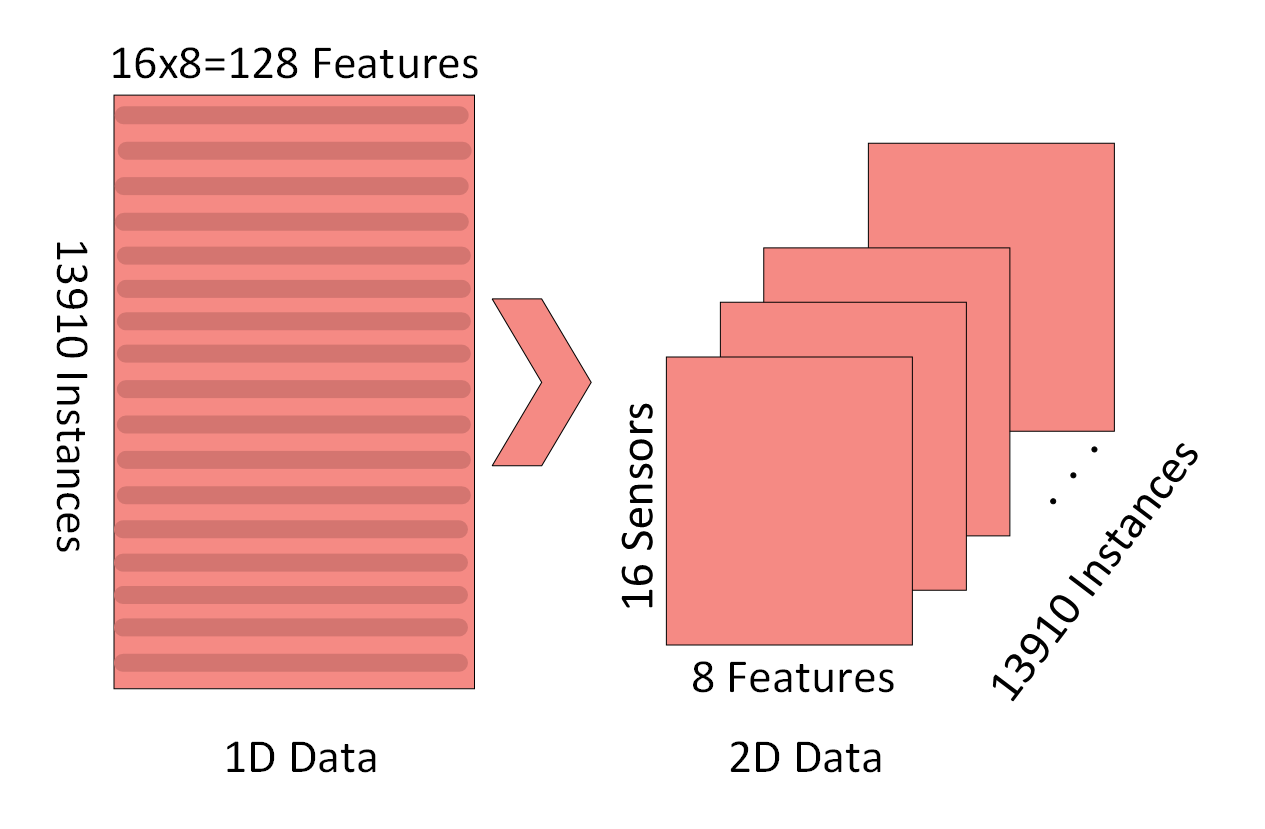}
      \caption{Data Conversion from 1D to 2D.}
      \label{CNN Data Rep}
      \vspace{-5mm}
   \end{figure}

In the artificial olfactory receptor module, feature extraction via CNN involves transforming the dataset into a 2D format while preserving sensor-specific feature details as depicted in Figure \ref{CNN Data Rep}. The dataset consists of 13910 instances, with each instance containing 128 features captured by 16 sensors. Specifically, each sensor records 8 distinct features. Thus, we have converted the initial 128 x 1D instance into an 8 x 16 2D instance, where 8 represents the feature set captured by each of the 16 sensors.
By converting gas sensor data into a 2D format, CNNs can effectively exploit spatial relationships between sensor readings, enabling the extraction of meaningful features at different levels of abstraction.

\section{RESULTS and DISCUSSION}
\begin{figure}[thpb]
      \centering

      \includegraphics[scale=0.35]{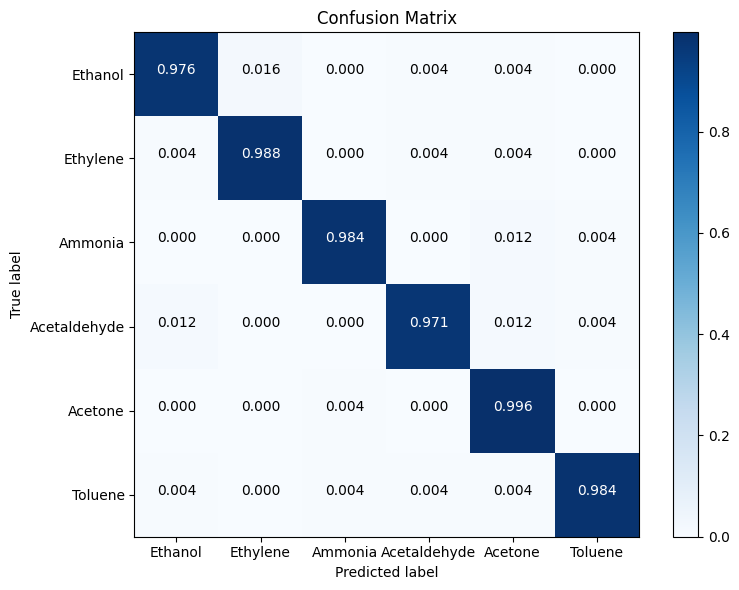}
      \caption{Model Classification Evaluation Results.}
      \label{confuse1}
      \vspace{-3mm}
   \end{figure}

The model's performance has been assessed across three distinct settings. In the first setting, the model's detection capability was evaluated by partitioning the entire dataset into training and testing subsets, with the achieved results illustrated in Fig \ref{confuse1}. A remarkable accuracy of 98.47\% was attained in this setting.

\begin{figure}[thpb]
      \centering

      \includegraphics[scale=0.28]{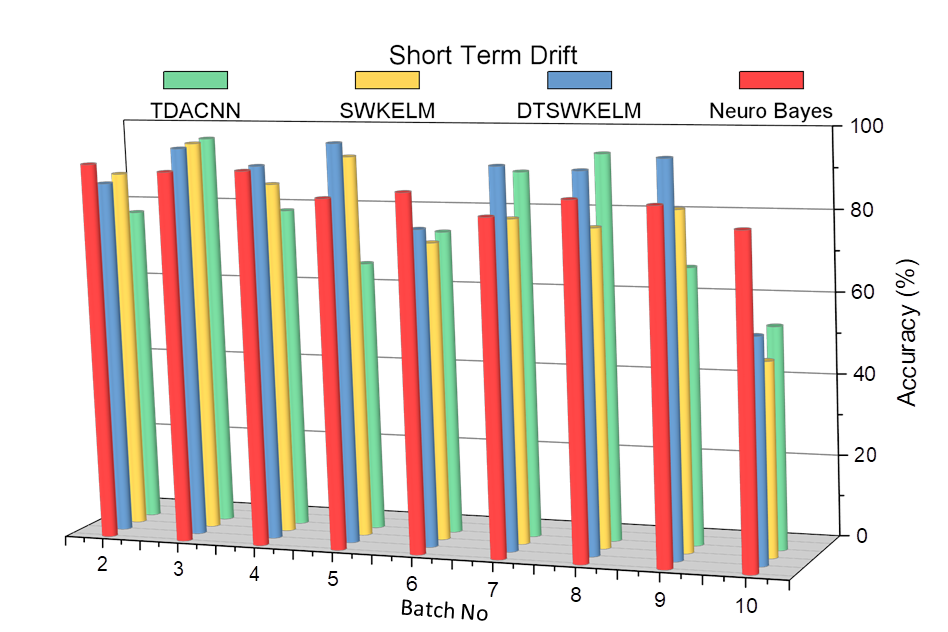}
      \caption{Short Term Drift  Response Comparison.}
      \label{Setting2}
      \vspace{-3mm}
   \end{figure}

Moving to the second setting, the model was trained on Batch(i) and subsequently tested on Batch(i+1). Results for this scenario are presented in Table \ref{Short Term} and Fig \ref{Setting2}. This setting assesses the model's behavior when subjected to short-term drift, yielding a satisfactory average accuracy of 88\%, showing the model's robustness.

\begin{figure}[thpb]
      \centering

      \includegraphics[scale=0.28]{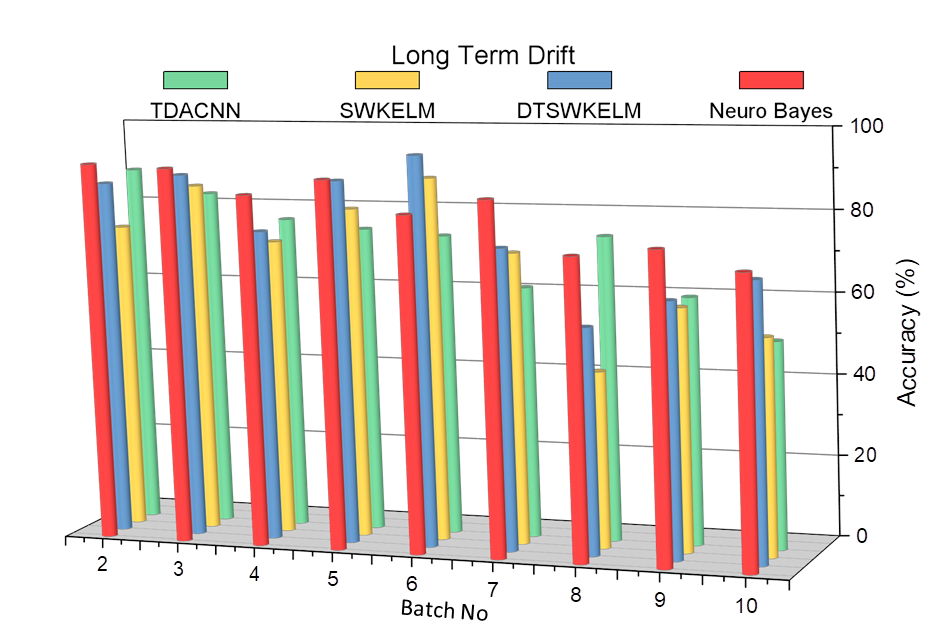}
      \caption{Long Term Drift  Response Comparison.}
      \label{Setting3}
      \vspace{-3mm}
   \end{figure}

In the third setting, the model underwent evaluation concerning long-term drift. Here, the model was trained on Batch 1 and tested on subsequent Batches from 2 to 10. Results are showcased in Table  \ref{Long Term} and Fig \ref{Setting3}. Comparisons with the results discussed in \cite{articleComp} are made, where \cite{articleComp} employed a domain transfer approach focusing on maximum mean discrepancy reduction based on extreme learning. The scores obtained from our Neuro-Bayesian model are provided at the end of both tables, demonstrating satisfactory performance relative to the comparison.

Notably, potential enhancements in performance could be attained by substituting the convolutional spiking neural network model with faithful replication of the human olfactory bulb, incorporating inhibitory feedback connections between granule cells and mitral cells. Additionally, implementing a certainty-based filtering mechanism could mitigate the impact of highly noisy and uncertain data entries on decision-making processes.

To evaluate the power consumption mitigation capability of the model, the mathematical model discussed in \cite{articlePower} is considered. When using 32-bit weight values, the energy consumption for a 32-bit integer Multiply and Accumulate ($E_{MAC}$) and only Accumulate ($E_{AC}$) operations is given in Table \ref{Energy Table}. The total inference energy (E) for an ANN or SNN, taking into account the total numbers of floating point operations (FLOPS) across all N layers of the network, is calculated as follows:


\begin{equation}
E_{ANN} =  (\sum_{l=1}^{N} FLOPS_{ANN})*E_{MAC} 
\end{equation}

\begin{equation}
E_{SNN} =  (\sum_{l=1}^{N} FLOPS_{SNN})*E_{AC}*T 
\end{equation}
where 'T' represents the number of time steps in the SNN model. For a single layer with output size O, kernel size k, input channels $C_{in}$ and output channels $C_{out}$ FLOPS calculations for ANN and SNN are given as:
\begin{equation}
FLOPS_{ANN} =  O^2*k^2*C_{in}*C_{out}
\end{equation}
\begin{equation}
FLOPS_{SNN} =  FLOPS_{ANN}*S_A
\end{equation}
$S_A$ is spiking activity, with average value of 0.23 in our model. The final calculations for energy efficiency are depicted in Table \ref{Total Energy Table}
\begin{table}[htbh]
\caption{Energy Consumption for 45nm CMOS Process}
 \label{Energy Table}
\begin{center}
\begin{tabular}{|p{2.5cm}|p{1.5cm}|}
\hline 
Operation 
 & Energy (pJ) 

\\
\hline
32-Bit Multiplication & 3.1
\\
\hline
32-Bit Addition & 0.1
\\
\hline
32-Bit MAC & 3.2 $E_{MAC}$ 
\\
\hline
32-Bit AC & 0.1 $E_{AC}$
\\
\hline

\end{tabular}
\end{center}
\vspace{-5mm}
\end{table}

\begin{table}[htbh]
\renewcommand{\arraystretch}{1}
\caption{Energy Calculation for both the Models}
 \label{Total Energy Table}
\begin{center}
\begin{tabular}{|p{1cm}|p{1.2cm}|p{1.5cm}|p{1.5cm}|p{1.4cm}|}
\hline 
Model
 & FLOPS & Parameters & Accuracy$_{avg}$ & Energy(nJ)  
\\
\hline
ANN
 & 482304 & 32K & 99 & 1543.4  
\\
\hline
SNN
 & 125254 & 32K & 98 & 626.3 
\\
\hline
\end{tabular}
\end{center}
\vspace{-10mm}
\end{table}

\begin{equation}
Energy Efficiency = \frac{E_{ANN}}{E_{SNN}} 
                = \frac{1543.4}{626.3}= 2.46              
\end{equation}
The above equation shows that the energy requirement for our model is twice less than that of a non-spiking model with the same number of layers and neurons.

\section{CONCLUSIONS} We've developed a model that emulates the functionality of the human olfaction system. Our model integrates spike-based Bayesian decision-making in an efficient way, a pivotal feature for addressing the generalization challenge in real-world environments. We rigorously tested the model's adaptability by subjecting it to a dataset containing drift-induced variations in gases, demonstrating its capacity to respond effectively in comparison to established benchmarks. Looking ahead, our future work envisions the implementation of an evolving Spiking Neural Network, incorporating feedback mechanisms within the artificial olfactory bulb, in conjunction with Bayesian inference. This holistic approach aims to enhance the model's performance, rendering it robust and application-agnostic, resilient to varying environmental conditions, and poised for broader utility.

\section*{ACKNOWLEDGMENT}

This work is supported by the Center for Autonomous Robotic Systems (KUCARS), Khalifa University, Abu Dhabi, UAE.






 \newpage
\bibliographystyle{IEEEtran}
\bibliography{IEEEabrv,Bibliography}

\end{document}